\documentclass[12pt]{iopart}

\begin{document}
\jl{1}
\title{Confinement effects from interacting chromo-magnetic and axion fields}
\author{Patricio Gaete\dag\footnote{e-mail address: patricio.gaete@usm.cl}, 
Euro Spallucci\ddag\footnote{e-mail address: spallucci@ts.infn.it}}
\address{\dag\ Departamento de F\'{\i}sica, Universidad T\'ecnica
F. Santa Mar\'{\i}a, Valpara\'{\i}so, Chile}
\address{\ddag\ Dipartimento di Fisica Teorica, Universit\`a di \
Trieste and INFN, Sezione di Trieste, Italy}

\begin{abstract}
We study a non-Abelian gauge theory with a pseudo scalar coupling
$\phi \, \mathrm{Tr}\left(\, F_{\mu \nu }^\ast
F^{\mu\nu}\,\right)$ in the case where a constant chromo-electric,
or chromo-magnetic, strength expectation value is present. We
compute the interaction potential within the framework of
gauge-invariant, path-dependent, variables formalism. While in the
case of a constant chromo-electric field strength expectation
value the static potential remains Coulombic, in the case of a
constant chromo-magnetic field strength the potential energy is
the sum of a Coulombic and a linear potentials, leading to the
\textit{confinement} of static charges.
\end{abstract}
\submitted
\maketitle
\section{Introduction}

It is presently widely accepted that quantitative understanding of
confinement of quarks and gluons remains as the major challenge in
$QCD$. In this connection, a linearly
 increasing quark-antiquark pair static potential provides the simplest
criterion for confinement, although unfortunately there is up to
now no known way to analytically derive
 the confining potential from first principles. In this context, it may be
recalled that phenomenological models have been of considerable
importance in order to provide strong
 insight into the physics of confinement, and can be considered as effective
theories of $QCD$. One of these, that is the dual
superconductivity picture of $QCD$
 vacuum \cite{Nambu}, has probably enjoyed the greatest popularity. The key
ingredient in this model is the condensation of topological
defects originated from quantum
 fluctuations (monopoles). As a consequence, the color electric flux linking
quarks is squeezed into ``strings'',  and the nonvanishing string
tension represents the proportionality
 constant in the linear, quark confining, potential. We further note that
recently an interesting approach to this problem has been proposed
\cite{Kondo}, which includes the
 contribution of all topologically  nontrivial sectors of a gauge theory.
Mention should be made, at this point,
 to lattice calculations which clearly show the formation of tubes of gluonic
fields connecting colored
 charges \cite{Capstick}.  In agreement with lattice  results,  loop-loop
correlations have been recently computed  analytically in
 extended  stochastic vacuum models  \cite{Shoshi}.\\
With these ideas in mind, in a previous paper \cite{GaeteG}, we
have studied a simple effective theory where confining potentials
are obtained in the presence of nontrivial constant expectation
values for the gauge field strength $F_{\mu \nu }$ coupled to a
scalar (axion) field $\phi$, via the interaction term
\begin{equation}
\mathcal{L}_I = \frac{g}{8}\,\phi\, \varepsilon ^{\mu \nu \alpha
\beta } F_{\mu \nu } F_{\alpha \beta }\ . \label{GS0}
\end{equation}
In particular, we have observed that in the case of a constant
electric field strength expectation value the static potential
remains Coulombic, while in the case of a constant magnetic field
strength expectation value the potential energy is the sum of a
Yukawa and a linear potential, leading to the confinement of
static charges. More interestingly, we remark that the magnetic
character of the field strength expectation value needed to obtain
confinement is in agreement  with the current chromo-magnetic
picture of the $QCD$ vacuum \cite{Savvidy}. Another feature of
this model is that it restores the rotational symmetry (in the
potential), despite of the external fields break this symmetry. We
further observe that similar results have been obtained in the
context of the dual Ginzburg-Landau theory \cite{Suganuma}, as
well as, for a theory of antisymmetric tensor fields that results
from the condensation of topological defects as a consequence of
the Julia-Toulouse mechanism \cite{GaeteW}. Accordingly, from a
phenomenological point of view, we have established an equivalence
between different models describing the same physical phenomena.
This allows us to obtain more information about a
theory than is possible by considering a single description.\\
By following this line of reasoning, it is natural to extend the
previous analysis to the case where a scalar field $\phi$  is
coupled to a \textit{non-Abelian} gauge field, that is,

\begin{equation}
{\mathcal L}_I  = \frac{\beta}{8}\, \phi\, \varepsilon ^{\mu \nu
\alpha \beta }\, \mathrm{Tr} \mathbf{F}_{\mu \nu } \mathbf{
F}_{\alpha \beta }\ .
 \label{GS1}
\end{equation}
where the trace is taken over color indices. 
While an Abelian model with interaction of the type (\ref{GS0}) is 
no more than a useful  ``laboratory'' ,  it may be worth
to recall that in $QCD$   coupling  of the form  (\ref{GS1})   is
instrumental to build up the Peccei -Quinn mechanism \cite{Peccei}
and solve the the strong $CP$ problem \cite{Kim,Wilczek,Weinberg}. 
In this case, the scalar field describes the axion,
i.e.  the Nambu-Goldstone boson of a new broken $U(1)$
symmetry of the quark and Higgs sector.\\   
The purpose of this work is to extend the Abelian calculations in
\cite{GaeteG} to the non-Abelian case and find the corresponding
static potential. Our calculations are done within the framework
of the gauge invariant/path-dependent variables formalism,
providing an effective tool for a better understanding of
effective non-Abelian theories. One important advantage of this
approach is that provides a physically-based alternative to the
usual Wilson loop approach, where in the latter the usual
qualitative picture of confinement in terms of an electric flux
tube linking quarks emerges naturally. As we shall see, in the
case of a constant chromo-electric field strength expectation
value the static potential remains Coulombic.  On the other hand,
in the case of a constant chromo-magnetic field strength
expectation value the potential energy is the sum of a Coulombic
and a linear potential, that is, the confinement between static
charges is obtained. As a result, the new coupling displays a
marked departure of a qualitative nature
from the results of Ref.\cite{GaeteG} at large distances.\\
It is interesting to observe that the dual superconductivity
picture of $QCD$, compared with the model proposed here, involves
the condensation of topological defects originated from quantum
fluctuations. Thus one is led to the conclusion that the
phenomenological model proposed here incorporates automatically
the contribution of the condensate of topological defects to the
vacuum of the model or, alternatively, the nontrivial topological
sectors as in \cite{Kondo}.  Another way of obtaining the above
conclusion is  by invoking the bosonization technique in
$(1+1)$-dimensions. Indeed, it is a well known fact that, for
instance, in the Schwinger model \cite{Schwinger}, the bosonized
version (effective theory) contains quantum corrections at the
classical level. In the same way, we can interpret the model
proposed here as an effective theory which contains quantum
effects at the classical level. Thus, one obtains a similarity
between the tree level mechanism that leads to confinement here
and the nonperturbative mechanism which gives confinement in
$QCD$. Accordingly, the above interrelations are interesting from
the point of view of providing unifications among diverse models
as well as exploiting the equivalence in explicit calculations, as
we are going to show.
\section{Interaction Energy}
As we discussed in the introduction, our immediate objective is to
calculate explicitly the interaction energy between static
pointlike sources, for a model containing the term (\ref{GS1}),
along the lines of Refs.\cite{GaeteG,Pato}. To this end we will
compute the expectation value $\left\langle H \right\rangle _\Phi$
of the Hamiltonian operator $H$ in the physical state $\left| \Phi
\right\rangle$ describing the sources. The non-Abelian gauge
theory we are considering is defined by the following generating
functional in four-dimensional spacetime:

\begin{equation}
{\mathcal Z} = \int {{\mathcal D}\phi {\mathcal D}A\exp \left\{ {
- i\int {d^4 x{\mathcal L}} } \right\}}\ , \label{GS2}
\end{equation}

with

\begin{equation}
{\mathcal L} =  - \frac{1}{4} \mathrm{Tr}\mathbf{F}_{\mu \nu
}\mathbf{F}^{\mu \nu }
 +\frac{\beta}{8}\,\phi\, \varepsilon ^{\mu \nu \rho \sigma } \,\mathrm{Tr}
\mathbf{F}_{\mu \nu }  \mathbf{F}_{\rho \sigma } +
\frac{1}{2}\partial_\mu \phi \, \,\partial^\mu \phi +\frac{{m_A^2
}}{2}\phi ^2\ , \label{GS3}
\end{equation}
where $m_A$ is the mass for the axion field $\phi$. Here,
$\mathbf{A}_\mu  \left( x \right) = A_\mu^a \left( x \right)T_a$,
where $T_a$ is a Hermitian representation of the semi-simple and
compact gauge group; and $F_{\mu \nu }^a  = \partial _\mu  A_\nu
^a  -
\partial _\nu  A_\mu ^a  + g\,
f^a_{bc} A_\mu ^b A_\nu ^c$, with $f^a_{bc}$ the structure
constants of the group. As in \cite{GaeteG}  we restrict ourselves
to static scalar fields, a consequence of this is that one may
replace $\partial^2\phi = - \nabla ^2\phi$. It also implies that, after
performing the integration over $\phi$ in ${\mathcal Z}$, the
effective Lagrangian density is given by
\begin{equation}
{\mathcal L} =  - \frac{1}{4}\,\mathrm{Tr}\mathbf{F}_{\mu\nu
}\mathbf{F}^{\mu \nu} + \frac{\beta^{2}}{128} \varepsilon^{\mu \nu
\rho \sigma }  \mathrm{Tr}\mathbf{F}_{\mu\nu}\mathbf{F}_{\rho
\sigma }\, \frac{1}{{\nabla ^2  - m_A^2 }} \varepsilon ^{\alpha
\beta \gamma\delta } \mathrm{Tr}\mathbf{F}_{\alpha \beta }
\mathbf{F}_{\gamma \delta }\ . \label{GS4}
\end{equation}

Furthermore, as was explained in \cite{Spallucci}, the expression
(\ref{GS4}) can be rewritten as

\begin{equation}
{\mathcal L} = - \frac{1}{4}\mathrm{Tr}\mathbf{f}_{\mu \nu
}\mathbf{f}^{\mu \nu }  + \frac{{\beta^2 }}{{16}}\varepsilon ^{\mu
\nu \alpha \beta } \mathrm{Tr}\left\langle {\mathbf{F}_{\mu\nu } }
\right\rangle \mathbf{f}_{\alpha \beta } \, \frac{1}{{\nabla ^2 -
m_A^2 }} \varepsilon ^{\rho \sigma \gamma \delta } \,
\mathrm{Tr}\left\langle {\mathbf{F}_{\rho \sigma } } \right\rangle
\mathbf{f}_{\gamma \delta }\ .  \label{GS5}
\end{equation}
where,  $\left\langle {{F}^a _{\mu \nu }} \right\rangle$
represents the constant classical background (which is a solution
of the classical equations of motion). Here, $f_{\mu \nu }^{a}$
describes a small fluctuation around the background, we also
mention that the above Lagrangian arose after using $\varepsilon
^{\mu \nu \alpha \beta }\mathrm{Tr} \left\langle \mathbf{F}_{\mu
\nu } \right\rangle\left\langle \mathbf{F}_{\alpha \beta }
\right\rangle  = 0$ (which holds for a pure chromo-electric or a
pure chromo-magnetic
background).\\
By introducing the notation $\varepsilon ^{\mu \nu \alpha \beta }
\left\langle{\mathbf{F}_{\mu \nu }} \right\rangle  \equiv
\mathbf{v}^{\alpha \beta }$ and $\varepsilon ^{\rho \sigma
\gamma\delta } \left\langle \mathbf{F}_{\rho\sigma } \right\rangle
\equiv \mathbf{v}^{\gamma \delta }$, expression (\ref{GS5}) then
becomes

\begin{equation}
{\mathcal L} =  - \frac{1}{4} \mathrm{Tr}\mathbf{f}_{\mu \nu }
\mathbf{f}^{\mu\nu } + \frac{{\beta^2 }}{{16}}
\mathrm{Tr}\mathbf{v}^{\alpha \beta } \mathbf{f}_{\alpha \beta }\,
\frac{1}{{\nabla ^2  - m_A^2 }} \,\mathrm{Tr}\mathbf{v}^{\gamma
\delta } \mathbf{f}_{\gamma \delta}\ . \label{GS6}
\end{equation}
At this stage we note that (\ref{GS6}) has the same form as the
corresponding Abelian effective Lagrangian density. This common
feature is our main motivation to study the effect of the
non-Abelian coupling on the interaction energy.

\subsection{Chromo-magnetic case.}
We now proceed to obtain the interaction energy in the $ {\bf
v}^{oi} \ne 0$ and ${\bf v}^{ij}=0$ case (referred to as the
chromo-magnetic one in what follows), by computing the expectation
value of the Hamiltonian in the physical state $\left| \Phi
\right\rangle$. The Lagrangian (\ref{GS6}) then becomes
\begin{equation}
{\mathcal L} =  - \frac{1}{4} \mathrm{Tr} \mathbf{f}_{\mu \nu }
\mathbf{f}^{\mu \nu}  + \frac{{\beta^2}}{{4}} \mathrm{Tr
}\mathbf{v}^{0i}  \mathbf{f}_{0i} \frac{1}{{ \nabla ^2  - m_A^2 }}
\mathrm{Tr} \mathbf{v}^{0k} \mathbf{f}_{0k} -
\mathrm{Tr}\mathbf{A}_0 \mathbf{J}^0  \label{GS7}
\end{equation}
where ${\bf J}^{0}$ is an external current, $(\mu ,\nu  =
0,1,2,3)$ and $(i,k= 1,2,3)$. Once this is done, the canonical
quantization in the manner of $Dirac$ yields the following
results. The canonical momenta are

\begin{eqnarray}
&&\Pi^{a0}=0, \label{GS8}\\
&& \Pi^{a} _i  = D^{ab}_{ij} E_b^{\,\, j}, \label{GS9}\\
&& E^{a}_i  \equiv f^{a}_{i0}\ ,\\
&& D^{ab}_{ij} \equiv \left( {\delta^{ab}\delta_{ij} +
\frac{{\beta^2 }}{2}v^{a}_{i0} \frac{1}{{ \nabla ^2  -
m_A^2}}v^{b}_{j0} } \right)\ .
\end{eqnarray}
Since $\mathbf{D}$ is a nonsingular,  there exists the inverse
$\mathbf{D}^{-1}$ and from Eq.(\ref{GS9}) we obtain
\begin{equation}
E^{a}_i  = \frac{1}{{\det \mathbf{D}}}\left\{
{\delta^{ab}\delta_{i}^j \det\mathbf{ D} -
\frac{{\beta^2}}{2}v^{a}_{i0} \frac{1}{{ \nabla ^2  -
m_A^2}}v^{b}_{j0} } \right\}\Pi^{b\, j} . \label{GS10}
\end{equation}

The corresponding canonical Hamiltonian is thus
\begin{eqnarray}
H_C &&= \int {d^3} x\, \mathrm{Tr}\left[\,  \mathbf{\Pi} ^i
\left(\, \mathbf{D} \, \mathbf{A}_0\,\right)_i +
\frac{1}{2}{\mathbf \Pi}^i
 {\mathbf \Pi}_i   +\frac{1}{2}{\mathbf B}^i {\mathbf B}_i
 \right.
 +\nonumber\\
&& \left. -\frac{\beta ^2 }{4}\left( {\mathbf v}^i {\mathbf \Pi}_i
\right)\frac{1}{\nabla ^2  - M^2 }\left(\, \mathbf{v}^i {\mathbf
\Pi} _i  \right) + \left(\, \mathbf{A}_0\mathbf{J}^0\,\right)
\right]\ ,  \label{GS11}
\end{eqnarray}
where $M^2  \equiv m_A^2  - \frac{{\beta^2 }}{8}{\mathbf v}^i
{\mathbf v}_i$  and ${\mathbf B}^i$ is the
\textit{chromo-magnetic} field. By applying Dirac  quantization
procedure for constrained systems, and removing   non-physical
variables by imposing an appropriate gauge condition \footnote{In
our gauge-invariant, path-dependent formalism, gauge fixing
procedure is equivalent to the choice of a particular path
\cite{Pato}, e.g.  a spacelike, straight, path $x^i= \xi^i +
\lambda\,\left(\, x -\xi \,\right)^i$, on a fixed time slice.}, we
can compute the interaction energy between pointlike sources in
the model under consideration. A fermion is localized at ${\mathbf
0}$ and an antifermion at $ {\mathbf y}$. From our above
discussion, we see that $\left\langle H \right\rangle _\Phi$ reads
\begin{equation}
\left\langle H \right\rangle _\Phi   = \left\langle \Phi
\right|\int {d^3} x \mathrm{Tr}\left[ \,
{\frac{1}{2}\mathrm{Tr}{\mathbf \Pi} ^i {\mathbf \Pi}_i -
\frac{{\beta ^2 }}{4}\mathrm{Tr} {\mathbf v}^i {\mathbf \Pi}_i  \,
\frac{1}{{\nabla ^2  - M^2 }}\mathrm{Tr}{{\mathbf v}^i {\mathbf
\Pi} _i } }\,
 \right]\left| \Phi  \right\rangle\ . \label{GS15}
\end{equation}
Now we recall that the physical state can be written as \cite{Pato},
\begin{equation}
\left| \Phi  \right\rangle  = \overline \psi  \left( {\mathbf y}
\right)P\exp \left( {ig\int_{\mathbf 0}^{\mathbf y} {dz^i
\mathbf{A}_i \left( z \right)} } \right) \psi \left( {\mathbf 0}
\right)\left| 0 \right\rangle\ .    \label{GS16}
\end{equation}

The line integral is along a spacelike path on a fixed time slice,
$P$ is the path-ordering prescription and $\left| 0\right\rangle$
is the physical vacuum state. As in \cite{Pato}, we again restrict
our attention to the weak coupling limit. From this and the
foregoing Hamiltonian discussion, we then get
\begin{equation}
\left\langle H \right\rangle _\Phi   = \left\langle H
\right\rangle _0  + V_1  + V_2, \label{GS18}
\end{equation}
where $\left\langle H \right\rangle _0  = \left\langle 0
\right|H\left| 0 \right\rangle$, and the $V_1$ and $V_2$ terms are
given by:
\begin{equation}
V_1  = \frac{1}{2}\left\langle \Phi  \right|\int
d^3x\,\mathrm{Tr}{\mathbf \Pi} ^i  {\mathbf \Pi} _i\,\left| \Phi
\right\rangle, \label{GS19}
\end{equation}
and
\begin{equation}
V_2  =  - \frac{\beta ^2 }{4}\left\langle \Phi  \right|\int {d^3 }
x \mathrm{Tr}{\mathbf v}^i {\mathbf \Pi} _i \, \frac{1}{\nabla ^2
- M^2 }\, \mathrm{Tr}{\mathbf v}^i  {\mathbf \Pi} _i \left| \Phi
\right\rangle\ . \label{GS20}
\end{equation}

One immediately sees that the $V_{1}$ term is identical to the
energy for the Yang-Mills theory. Notwithstanding, in order to put
our discussion into context it is useful to summarize the relevant
aspects of the analysis described previously \cite{Pato}.  Thus,
we then get an Abelian part (proportional to $C_{F}$) and a
non-Abelian part (proportional to the combination $C_{F}C_{A}$).
As we have noted before, the Abelian part takes the form
\begin{equation}
V^{(g^2) }  = \frac{1}{2}g^2 \mathrm{Tr}(T^a T_a)\, \int_{\mathbf
0}^{\mathbf y} {dz^i } \int_{\mathbf 0}^{\mathbf y} {dz_i } \,
\delta \left( {{\mathbf z} - {\mathbf z}^ \prime  } \right)\ ,
\label{GS21}
\end{equation}
remembering that the integrals over $z^i$ and $z_i^{\prime}$ are
zero except on the contour of integration. Writing the group
factor $\mathrm{Tr}T^{a}T_{a}=C_{F}$, the expression (\ref{GS21})
is given by
\begin{equation}
V^{(g^2) } \left( L \right) =  - \frac{g^2C_F}{{4\pi \, L}} \ ,
\label{GS22}
\end{equation}
where $\left| {\mathbf y} \right| \equiv L$. Next, the non-Abelian
part may be written as
\begin{equation}
V^{(g^4) }  = \mathrm{Tr}\int {d^3 } x\left\langle 0 \right|\,
\mathbf{I}^i  \mathbf{I}_i \left| 0 \right\rangle\ , \label{GS23}
\end{equation}
where,
\begin{equation}
I^{a\, i } = g^2 f^a_{bc} T^b \int_{\mathbf 0}^{\mathbf y} {dz^k }
\int_0^1 {d\lambda } A_k^c \left( z \right)z^i \delta \left(
{{\mathbf x} - \lambda {\mathbf z}} \right). \label{GS24}
\end{equation}
It should be noted that, by using spherical coordinates,
expression (\ref{GS24}) reduces to
\begin{equation}
 I^{a\, i } = g^2 f^a_{bc}  T^b \frac{{{\mathbf z}^i }}{{\left| {\mathbf z}
\right|}} \frac{1}{{\left| {\mathbf x} \right|^2 }}\int_{\mathbf
0}^{\mathbf y} {dz^k } A_k^c \left( z \right)\sum\limits_{lm}
{Y_{lm}^\ast  } \left( {{\theta}^ \prime  ,{\varphi}^\prime }
\right)Y_{lm} \left( {\theta ,\varphi } \right)\ . \label{GS25}
\end{equation}
Putting this back into Eq.(\ref{GS23}), we obtain

\begin{equation}
V^{g^4 } \left(\, L\, \right) = -C_A C_F \left( \frac{ g^4 }{2L}
\right)\int_{\mathbf 0}^{\mathbf y} {dz^i } \int_{\mathbf
0}^{\mathbf y} {dz^{ \prime j} } D_{ij} \left( {z,z^ \prime  }
\right)\ . \label{GS26}
\end{equation}
Here $D_{ij}(z,z^{\prime})$ stands for the propagator, wich is
diagonal in color and taken in an arbitrary gauge. Following our
earlier discussion, we choose the Feynman gauge. As a consequence,
expression (\ref{GS26}) then becomes
\begin{equation}
V^{g^4 } \left(\, L\, \right) =  - g^4 \frac{1}{{4\pi ^2 }}C_A C_F
\frac{1}{L}\log \left( {\Lambda L} \right)\ , \label{GS27}
\end{equation}
where $\Lambda$ is a cutoff. Then, the $V_{1}$ term takes the form
\begin{equation}
V_1  =  - g^2 C_F \frac{1}{{4\pi L}} \left( {1 + \frac{{g^2 }}{\pi
}C_A \log \left( {\Lambda L} \right)} \right)\ . \label{GS28}
\end{equation}
It is important to realize that our calculation was based only on
the Hamiltonian and on the geometrical requirement that the
fermion-antifermion state be invariant under gauge
transformations. From (\ref{GS28}) we see that the term of order
$g^{2}$ is just the Coulomb energy due to the color charges of the
quarks. The correction term of order $g^{4}$ represents an
increase of the energy due to the vacuum fluctuations of the gauge
fields.

The task is now to evaluate the $V_{2}$ term, which is given by
\begin{equation}
V_2  =  - \frac{{\beta ^2 }}{4}\,\left\langle \Phi  \right|\int
{d^3 } x \mathrm{Tr}{{\mathbf v}^i  {\mathbf \Pi} _i }
\frac{1}{{\nabla ^2  - M^2 }} \mathrm{Tr}{{\mathbf v}^j {\mathbf
\Pi} _j } \left| \Phi  \right\rangle\ . \label{GS29}
\end{equation}
Once again, from our above Hamiltonian structure we have an
Abelian contribution and a non-Abelian contribution, in other
words,
\begin{eqnarray}
V_2^{(Ab.)}  &&=  - \frac{{\beta ^2 }}{4}g^2
\mathrm{Tr}\mathbf{v}^i \mathbf{v}_i  \, \int {d^3 }
x\int_{\mathbf 0}^{\mathbf y} {dz ^{\prime i} } \delta \left(
{{\mathbf x} - {\mathbf z}^{\prime}  } \right)\, {\frac{1}{{\nabla
_x^2  - M^2 }}} \,\int_{\mathbf 0}^{\mathbf y} {dz^i } \delta
\left( {{\mathbf x}
- {\mathbf z}} \right)\ ,\nonumber\\
&&  \label{GS30}
\end{eqnarray}
and
\begin{equation}
V_2^{\left( {Non-Ab.} \right)}  = \frac{\beta ^2g^4}{4}
\mathrm{Tr}\left( \, T^bT^d \, \right) f_{pbc}f_{qd}^c  v^{pi}
v^q_i \int_{\mathbf 0}^{\mathbf y} {dz^l } \int_{\mathbf
0}^{\mathbf y} {dz^{ \prime k} } D_{lk} \left( {{\mathbf
z},{\mathbf z}^ \prime } \right)\int_{\mathbf 0}^{{\mathbf z}^
\prime  } {du^j } \int_{\mathbf 0}^{\mathbf z} {dv_j } G\left(
{{\mathbf u},{\mathbf v}} \right)\ , \label{GS31}
\end{equation}
as before, $D_{lk}(z,z^{\prime})$ represents the propagator. Here,
$G$ is the Green function
\begin{equation}
G\left( {{\mathbf u},{\mathbf v}} \right) = \frac{1}{{4\pi
}}\frac{{e^{ - M\left| {{\mathbf z}^ \prime  - {\mathbf z}}
\right|} }}{{\left| {{\mathbf z}^ \prime
 - {\mathbf z}}
\right|}}\ .\label{GS32}
\end{equation}
This Green function is, in momentum space,
\begin{equation}
\frac{1}{{4\pi }}\frac{{e^{ - M|{\mathbf u}   - {\mathbf v}|}
}}{{|{\mathbf u} - {\mathbf v}|}} = \int {\frac{{d^3 k}}{{\left(
{2\pi } \right)^3 }}\frac{{e^{i{\mathbf k} \cdot \left( {{\mathbf
u}- {\mathbf v}} \right)} }}{{{\mathbf k}^2  + M^2 }}}\ .
\label{GS33}
\end{equation}
By means of Eq. (\ref{GS33}) and remembering that the integrals
over $z^{i}$ and $z^{\prime i}$ are zero except on the contour of
integration, the term (\ref{GS30}) reduces to the linearly
increasing potential \cite{GaeteG}, that is,

\begin{equation}
V_2^{\left( {Ab.} \right)}  = \frac{{\beta ^2 g^2 }}{{16\pi }}
\mathrm{Tr}\left( \, \mathbf{v}^i  \mathbf{v}_i \,  \right)\, L\,
\log \left( {1 + \frac{{\Lambda ^2 }}{{M^2 }}} \right)\ ,
\label{GS34}
\end{equation}

We now proceed to calculate the $V_2^{\left( {Non-Ab.} \right)}$
term. As before, we will use the Green function (\ref{GS33}) in
momentum space to handle the integral in Eq.(\ref{GS31}).
Following our earlier procedure \cite{GaeteG}, Eq.(\ref{GS31}) is
further rewritten as

\begin{eqnarray}
V_2^{\left( {Non-Ab.} \right)}  =&& \frac{{\beta ^2 g^4 }}{8}
\mathrm{Tr}\left(\,T^bT^d\,\right)
f_{pbc}f_{qd}^cv^{pi}v^q_i
\times\nonumber\\
&&\log \left( \, 1 + \frac{\Lambda ^2 }{M^2 } \right)\int_{\mathbf
0}^{\mathbf y} {dz^l } \int_{\mathbf 0}^{\mathbf y} {dz^{ \prime
k} } \left| {\mathbf z} \right|D_{lk} \left( {{\mathbf z}\
,{\mathbf z}^ \prime  } \right)\ . \label{GS35}
\end{eqnarray}

Now, we move on to compute the integral (\ref{GS35}). As in the
previous calculation, we choose $D_{lk} ({\mathbf z},{\mathbf z}^
\prime  )$ in the Feynman gauge. Thus the  $V_2^{\left(
{Non-Abelian} \right)}$ term is
\begin{equation}
V_2^{\left( {Non-Ab.} \right)}  = \frac{{\beta ^2 g^4
}}{8}\mathrm{Tr}\left( \,T^b  T^d\, \right)
f^{pbc}f_{qd}^c v^{pi}
v^q_i L\, \log \left( {1 + \frac{{\Lambda ^2 }}{{M^2 }}} \right),
\label{GS36}
\end{equation}
after substracting the self-energy terms.

From (\ref{GS34}) and (\ref{GS36}) we then get
\begin{equation}
V_2  = \frac{\beta ^2 g^2 }{8}\left[\,
 \frac{1}{2\pi } \mathrm{Tr}\left( \, \mathbf{v}^i  \mathbf{v}_i \,  \right)\,
+ g^2   \mathrm{Tr}\left( \, T^b  T^d  \, \right)\,
f_{pbc}f_{qd}^c v^{pi}v^q_i  \, \right]\, L\log \left(\, 1 +
\frac{{\Lambda ^2 }}{{M^2 }}\,\right)\ . \label{GS37}
\end{equation}

By putting together Eqs. (\ref{GS28}) and (\ref{GS37}), we obtain
for the total interquark potential
\begin{eqnarray}
V= &&-  \frac{ g^2 C_F }{4\pi L} \left[ \,  1 + \frac{g^2 }{\pi}C_A
\log \left( \, \Lambda L\, \right) \, \right]+\nonumber\\
&& + \frac{\beta ^2 g^2 }{8}  \,   \left[ \,\frac{1}{2\pi }
 \mathrm{Tr}(\mathbf{v}^i  \mathbf{v}_i)   +
g^2  \,  \mathrm{Tr}\left( \, T^b  T^d \, \right) \, f_{pbc}
f_{qd}^c v^{pi}v^q_i  \, \right] \, L\log \, \left(\, 1 +
\frac{\Lambda ^2 }{M^2
}\,\right)\ . \nonumber\\
&& \label{GS38}
\end{eqnarray}

It must be observed that the rotational symmetry is restored in
the resulting form of the potential, although the external fields
break the isotropy of the problem in a manifest way. It should be
remarked that this feature is also shared by the corresponding
Abelian interaction energy \cite{GaeteG}. As we have noted before
this improves the situation as compared to the "spaghetti vacuum"
model \cite{Savvidy} where rotational
symmetry seem to be very difficult to restore. \\
Now we recall the calculation reported in \cite{Kondo} by taking
into account topological nontrivial sectors in $U(1)$ gauge theory
\begin{equation}
V\left( L \right) =  - \frac{e^2 }{4\pi }\frac{1}{L} + \sigma
L\ . \label{GS39}
\end{equation}
We immediately see that the result (\ref{GS38}) is exactly the one
obtained by Ref.\cite{Kondo}. It is
 interesting to notice that even if (\ref{GS5}) is an effective model, extending
 the Abelian one discussed in \cite{GaeteG}, it is able to reproduce the 
 correct form of the Cornell potential (\ref{GS39}). As such it deserves some 
 further investigation.

\subsection{Chromo-electric case.}
Now we focus on the case ${\bf v}^{0i}=0$ and ${\bf v}^{ij}\ne 0$
(referred to as the chromo-electric one in what follows). The
corresponding Lagrangian density reads
\begin{equation}
\mathcal{ L} =  - \frac{1}{4} \mathrm{Tr} \mathbf{f}_{\mu
\nu}\mathbf{f}^{\mu \nu }  + \frac{\beta^2}{16}
\mathrm{Tr}\mathbf{v}^{ij} \mathbf{f}_{ij} \frac{1}{ \nabla ^2  -
m_A^2 }\mathrm{Tr}\mathbf{v}^{kl} \mathbf{f}_{kl}
 -\mathrm{Tr}\mathbf{A}_0 \mathbf{J}^0\ , \label{GS40}
\end{equation}
$(\mu ,\nu  = 0,1,2,3)$ and $(i,j,k,l = 1,2,3)$.\\
Here again, the quantization is carried out using  Dirac's
procedure. We can thus write the canonical momenta
$\Pi^{a\mu}=-f^{a0\mu}$, which results in the usual primary
constraint $\Pi^{a\, 0}=0$ and $\Pi^{a\, i}=f^{a\, i0}$. Defining
the electric and magnetic fields by $ E^{a\, i}  = f^{i0}$ and
$B^{a\, k}  = -\frac{1}{2}\varepsilon ^{kij} f^{a\, ij}$,
respectively, the canonical Hamiltonian is thus

\begin{eqnarray}
H_C  &&= \int {d^3 } x  \left[ \,
\frac{1}{2}\mathrm{Tr}\mathbf{E}^i \mathbf{ E}_i +
\frac{1}{2}\mathrm{Tr}\mathbf{B}^i {\mathbf B}_i
-\frac{\beta^2}{16} \varepsilon_{ijm}\varepsilon_{kln}
\mathrm{Tr}\mathbf{v}^{ij}\mathbf{B}^m \,\frac{1}{{ \nabla ^2  -
m_A^2 }}   \mathrm{Tr}
\mathbf{v}^{kl}\mathbf{B}^n  \right.\nonumber\\
&& \left. -\mathrm{Tr}\mathbf{A}_0 \left(\, \partial _i
\mathbf{\Pi} ^i - \mathbf{J}^{0} \, \right)\, \right]\ .
\label{GS41}
\end{eqnarray}
 It is straightforward to see that the
constrained structure for the gauge field is identical to the
usual Yang-Mills theory. However, in order to put the discussion
into the context of this paper, it is convenient to mention the
relevant aspects of the analysis described previously \cite{Pato}.
Therefore, we pass now to the calculation of the interaction energy.\\
As done above, our objective is now to calculate the expectation
value of the Hamiltonian in the physical state $\left| \Phi
\right\rangle$. In other words,
\begin{equation}
 \left\langle H \right\rangle _\Phi   =  \left\langle \Phi
\right| \,\frac{1}{2} \int {d^3 x} \, \mathrm{Tr}{\mathbf
E}^i{\mathbf E}_i \, \left|\Phi \right\rangle\ . \label{GS42}
\end{equation}
Taking into account the above Hamiltonian structure, the
interaction takes the form
\begin{equation}
\langle H\rangle _{\Phi }=\langle H\rangle _{0}+V_{1},
\label{GS43}
\end{equation}
where $\langle H\rangle _{0}=\langle 0\mid H\mid 0\rangle$.
Accordingly, the potential reads
\begin{equation}
V_1  =  - \frac{  g^2 C_F }{{4\pi L}}\left( {1 + \frac{{g^2 }}{\pi
}C_A \log \left( {\Lambda L} \right)} \right)\ . \label{GS44}
\end{equation}

\section{Final Remarks}

In summary, we have considered the confinement versus screening
issue for a non-Abelian theory with a coupling $\varepsilon ^{\mu
\nu \alpha \beta } F_{\mu \nu }^a F_{\alpha \beta }^a$, in the
case when there are nontrivial constant expectation values for the
gauge field strength $F^a_{\mu \nu }$. The constant gauge field
configuration is a
solution of the classical equations of motion.\\
It was shown that in the case when $\left\langle {F^{a}_{\mu \nu }
} \right\rangle$ is chromo-electric-like no unexpected features
are found. Indeed, the resulting static potential remains
Coulombic. More interestingly, it was shown that when
$\left\langle {F^{a}_{\mu \nu } }\right\rangle$ is
chromo-magnetic-like the potential between static charges displays
a Coulomb piece plus a linear confining piece. An analogous
situation in the Abelian case may be recalled \cite{GaeteG}. Also,
a common feature of these models (Abelian and non-Abelian) is that
the rotational symmetry is restored in the resulting interaction
energy.\\   We recall that the effective action (\ref{GS6}) is the
expansion of (\ref{GS5}) up to second order in the fluctuation
field  $\mathbf{f}_{\mu\nu}$, thus one could wonder about the
effect of including higher order terms. Even if an explicit
calculation is well beyond the purpose of this note, some general
comments can be done. We know that higher order quantum effects
renormalize coupling constants by making them scale dependent. It
is easy to see that the logarithmic correction to the Coulombic
term in (\ref{GS38}) is nothing but the first order expansion of
the effective, running, coupling constant

\begin{equation}
g^2_{eff}\left(\, \Lambda L \, \right)= \frac{g^2\pi}{ 1 -
\frac{g^2 C_A}{\pi } \log \left(\, \Lambda L \,\right)},
\label{GS45}
\end{equation}

$g^2_{eff}$ exhibits the expected \textit{asymptotically free}
behavior at short distance characterizing the non-Abelian
character of the strong interaction.  This result marks a clear
difference from the Abelian case, where the electric charge
increases at short distance. Similar renormalization effects are
expected for the string tension $\sigma$,  as well. However, the $
log( 1 + \frac{\Lambda^2} {M^2} )$ factor in the second term in
(\ref{GS38}) is L-independent. We are confident that higher order
corrections will preserve this feature leading to a still
confining static potential of the form

\begin{equation}
V(L) =  - \frac{{g_{eff}^2 \left( {\Lambda L} \right)C_F }}{{4L}}
+ \sigma _{eff} \left(\, {\Lambda ^2 /M^2 } \, \right)L \label{GS46}
\end{equation}

An explicit check of  (\ref{GS46}) will be addressed
in a future work.\\
We conclude by noting that our result agrees with the monopole
plasma mechanism \cite{Nambu}, \cite{Kondo}. However, although
both approaches lead to confinement, the above analysis reveals
that the mechanism of obtaining a linear potential is quite
different. As already mentioned, in this work we have exploited
the similarity between the tree level mechanism that leads to
confinement here and the nonperturbative mechanism (caused by
monopoles) which gives confinement in $QCD$.

\section{ACKNOWLEDGMENTS}

One of us (P.G.) wants to thank the Physics Department of the
Universit\`a di Trieste for hospitality and I.N.F.N. for support.
Work supported in part by Fondecyt (Chile) grant 1050546 (P.G.).

\end{document}